\documentclass[pra,twocolumn,showpacs,floatfix,aps]{revtex4-1}

\usepackage{amsmath}

\usepackage{graphicx}
\newcommand{\erf}{{\rm erf}}
\newcommand{\dr}{{\rm d}}
\newcommand{\rvec}{\boldsymbol{r}}

\begin{document}

\title{Implementation strategies for multiband quantum simulators of real materials}
\author{J.P. Hague} \affiliation{School of Physical Sciences, The
  Open University, MK7 6AA, UK}

\author{C.MacCormick} 
\affiliation{School of Physical Sciences, The Open University, MK7 6AA, UK}

\begin{abstract}
The majority of quantum simulators treat simplified one-band strongly correlated models, whereas multiple bands are needed to describe materials with intermediate correlation. We investigate the sensitivity of multiband quantum simulators to: (1) the form of optical lattices (2) the interactions between electron analogues. Since the kinetic energy terms of electron analogues in a quantum simulator and electrons in a solid are identical, by examining both periodic potential and interaction we explore the full problem of many-band quantum simulators within the Born-Oppenheimer approximation. Density functional calculations show that bandstructure is highly sensitive to the form of optical lattice, and it is necessary to go beyond sinusoidal potentials to ensure that the bands closest to the Fermi surface are similar to those in real materials. Analysis of several electron analogue types finds that dressed Rydberg atoms (DRAs) have promising interactions for multi band quantum simulation.  DRA properties can be chosen so that interaction matrices approximate those in real systems and decoherence effects are controlled, albeit with parameters at the edge of currently available technology. We conclude that multiband quantum simulators implemented using the principles established here could provide insight into the complex processes in real materials.
\end{abstract}

\date{January 10, 2017}


\maketitle

\section{Introduction} 

Systems of cold atoms  have
been highly successful as quantum simulators for simplified one-band
models of condensed matter, such as the Hubbard model, where atoms move in sinusoidal optical lattices \cite{RevModPhys80885}. 
 Real
materials, on the other hand, consist of multiple electronic bands,
interacting via Coulomb repulsion and moving in a periodic lattice potential with form $\sum_i -k_e e^2 Z/|\rvec-\rvec_i|$, where $Z$ is the atomic number, $\rvec$ are the locations of electrons, $\rvec_i$ are the locations of atomic nuclei and $k_e=1/4\pi\epsilon_0$. It is therefore of interest to examine to what extent such a scenario could be reproduced with quantum simulators. Painted potentials and holograms could be used to make more realistic optical lattices from the condensed matter perspective. However, it is not possible to reproduce the precise form of the nuclear potential due to optical resolution effects, and this is likely to affect the band structure. Another issue is the interaction potential, which needs to be of a specific size and form. The relative energies of the bands could sensitively depend on the form of the external potential and magnitude of interactions between the atoms in the lattice. Often states far from the Fermi surface are less important, so simulation of solid state phenomena requires as a minimum that the bands near the Fermi surface are realistically represented. 

Since optical lattices are fixed, they naturally neglect phonon effects and are therefore exactly described within the Born--Oppenheimer approximation (BOA). The full many-body Hamiltonian of a solid state system within the BOA contains only three terms: Kinetic energy of the electrons, interaction potential between electrons and lattice and the interaction between electrons. Since the kinetic energy terms of electrons and atoms are the same, we can carry out a complete examination of implementation strategies for quantum simulators of many-body condensed matter Hamiltonians within the BOA by investigating two key aspects:
\begin{enumerate}
\item  Establishing to what extent fundamental limitations in the form of optical lattices affects the band structure of quantum simulators
that emulate the multi-band physics of materials. To the best of our knowledge this is the first time such an analysis has been carried out. We examine the use of painted potentials or holograms to form optical lattices that are good approximations to the periodic lattice potential. This can be achieved by convolving the $-k_e e^2 Z / r$ potential (subject to an energy scaling) with a Gaussian beam, such that when the beam waist is small compared to the optical lattice spacing, the $-k_e e^2 Z / r$ potential naturally emerges and when beam waist is large the sinusoidal lattice is recovered.
  
\item Identifying the best electron analogues to use in such optical lattice systems. None of the available electron analogues is ideal. We describe the strengths and limitations of systems of cold ions, dressed Rydberg atoms, polar molecules and cold atoms interacting via Feshbach resonance for approximating the interaction between electrons in condensed matter systems. We also discuss temperature constraints and decoherence.
\end{enumerate}

This paper uses graphene and BN as toy systems with well understood electronic structures that can be used to assess the proposed quantum simulation schemes. However, the general principles established here can be used to guide the implementation of quantum simulators for more complex materials. This would provide clean and tuneable systems to help understand the physics of condensed matter systems in a controlled way. Also, schemes following these principles could be used as much needed test beds for condensed matter numerics. 

Before continuing, we briefly describe the various limiting behaviors of condensed matter systems, where the ratio of the width of electronic bands to Coulomb interactions gives a rough
estimate of the level of quantum correlation. Typical cold atom quantum simulators to date have concentrated on the strongly correlated scenario of the Hubbard model, which while very interesting from the point of view of experimental development, is already well studied using numerical techniques, and convergence between these techniques is emerging \cite{leblanc2015}.
In this strong correlation limit, i.e. in materials where the electrons are well localized to
atoms, and electronic bands are very narrow, the Hubbard model is
accurate. Interaction in this limit is characterized by a single parameter, $U$. The weak $U$ limit is often probed in both optical lattice experiments and in numerical simulations.  However, from the point of view of condensed matter systems, the single-band Hubbard model is strictly invalid for any real system if $U$ is not strong relative to the electron hopping
between lattice sites, $t$ \cite{hubbard1963a} (perhaps with the exception of hydrogen under high pressure). Notwithstanding, there is some very interesting physics associated with the Hubbard model, including the Mott metal-insulator transition \cite{hubbard1964} (superfluid-insulator transition if bosons are used \cite{jaksch1998}), magnetism \cite{hubbard1965} and potentially superconductivity (see e.g. \cite{bickers1989}).

The single-band Hubbard model viewpoint is likely to miss much of the rich physics found in real condensed matter systems. The relative spacing and width of the bands is critically important to the size of interband interactions and in many systems the number of interacting bands is large.  For systems where the band width is small relative to Coulomb repulsion, multiple Hubbard parameters can be introduced to maintain accuracy (see e.g. \cite{gillan1973}). In systems where bands are broad compared to interaction energies, correlations are weak and Hubbard like approximations break down. The physics in the weak correlation limit is quite different and ab-initio techniques such as density functional theory (DFT) are
accurate and popular \cite{kohn1965a}. 

In the parameter space lying between weakly correlated systems and multi-band Hubbard models, there is a region of intermediate correlation where similar energy scales are associated with the width of electronic bands and the size of the Coulomb repulsion between electrons. This regime is difficult to simulate numerically to a good level of accuracy, and quantum simulators that can probe the physics of such intermediate correlation in systems with multiple bands (at least to some degree of approximation) could offer several benefits:
\begin{enumerate}
\item Cold atom quantum simulators that have multiple electronic bands could be used to examine intermediate correlation physics in a controllable way that would be useful as a benchmark for numerical methods, potentially leading to improvements in such approaches.
\item While detailed predictions might be out of the reach of quantum simulators, determining the best possible form and magnitude for the interactions could lead to better probes for the rich physics of condensed matter systems. 
\item Well separated core states are important for ensuring that the states near to the Fermi surface have the correct symmetries, since all states in a 1 band optical lattice have an $s$ character, whereas in condensed matter systems they may have $p$, $d$ or $f$ symmetry.
\item Also, interband interactions between itinerant and core states can be very important in some cases, such as the heavy fermion materials. 
\end{enumerate}

Typical experimental implementations of cold atom quantum simulators
have focused on the Hubbard model \cite{bloch2012a,Schneider05122008,Nature455,PhysRevLett94080403},
and other simulators made from cold ions or Rydberg atoms have been
constructed to examine spin systems
\cite{blatt2012a,PhysRevLett103035303,Greif14062013}. In
these systems, atoms or ions in the simulator represent electrons in a
simplified material, and an optical lattice represents the external
potential from a periodic array of nuclei  \cite{RevModPhys80885}. A wide range of theoretical
studies have focused on the strong correlation limits \cite{NaturePhysInsightCirac,PhysRevLett103035303,RevModPhys80885}. We also note that quantum simulators of the Hubbard and other models can be constructed using cold ions, by using the resulting spin systems in combination with the Jordan--Wigner transformation \cite{stojanovic2012a,casanova2012a}.

On the other hand, the weak correlation limit has had limited attention \cite{PhysRevLett112015301}. Of particular interest is the application of density functional theory to the band structures of
hydrogen like cold atom systems interacting via $\delta$-function like
potentials in a shallow sinusoidal optical lattice, simulated by Ma
{\it et al.} \cite{ma2012a}. The all atom (all electron) methods used
here and in Ref. \cite{ma2012a} work in continuous space, and should
not be confused with methods such as Bethe-ansatz local density approximation (BALDA) which
focus on Hubbard models on discrete lattices
\cite{xianlong2006a}. Before continuing, we wish to make it clear that we are not simulating Hubbard models in this paper, rather we are applying density functional theory techniques to cold atoms moving in continuous lattices.

There have been a range of studies where optical lattice experiments have moved away from simple square and cubic lattices to investigate more specialized systems. For example, honeycomb lattices typical to graphene have been implemented \cite{tarruell2012,duca2015}. The difference between these approaches, and the system proposed here is that the former simulate a single correlated band, whereas we propose quantum simulators that can treat all bands of a multiband system, including core and itinerant bands, in a single simulator. This is far closer to the physics of a real solid, and would make it possible to study how tuning the interactions in a material change its properties. Other benefits over single band quantum simulators were discussed above. We use graphene and BN here because they are well understood systems, but the aim is that the ideas presented here can be applied to other, more complex, materials.

This paper is structured as follows. In Sec. \ref{sec:modelmethod} we discuss the full electron Hamiltonian of a condensed matter system, and the considerations for simulating it using an optical lattice system. Ideally an optical lattice for condensed matter systems would faithfully reproduce the periodic $-k_e e^2 Z / r$ potentials due to the lattice, which imposes stringent conditions upon the design of the optical system. We also describe the density functional approach used to study the formation of energy bands in  systems with an approximate lattice potential due to an experimentally realizable optical system. In Sec. \ref{sec:resolution} we study specific simulated graphene and boron nitride monolayer systems. In Sec. \ref{sec:analogues} we discuss the magnitude and the form of interactions between different electron analogues, with the aim of establishing the best system to use in experimental implementations. In Sec. \ref{sec:coherence} we discuss operating temperatures and decoherence effects. Finally we summarize and conclude in Sec. \ref{sec:conclusions}.

\section{Model and method}
\label{sec:modelmethod}

In this section, we discuss our model for optical lattices formed by painted potentials and holograms and the method we use to investigate it. We start by discussing interactions in a typical condensed matter system.

The well-known full electron
Schr\"odinger equation for a condensed matter system has the form,
\begin{equation}
 -\frac{\hbar^2}{2m_e}\sum_{i}\nabla_{i}^2\psi + \sum_{i\alpha}V_{\rm ext}(\rvec_{i\alpha})\psi + \sum_{i<j}V_{\rm int}(\rvec_{ij})\psi= E_{\rm CM} \psi
\label{eqn:electronhamiltonian}
\end{equation}
where the wavefunction $\psi\equiv\psi(\rvec_i,\cdots,\rvec_n)$ is a many-body wavefunction depending on all coordinates of $n$ electrons, displacement between different electrons is $\rvec_{ij}\equiv \rvec_j-\rvec_i$, and $\rvec_{i\alpha}=\rvec_{\alpha}-\rvec_{i}$ is the displacement between electrons and nuclei. In the condensed matter system $V_{\rm ext}(\rvec_{i\alpha})=-k_e e^2 Z / r_{i\alpha}$, $m_{e}$ is the mass of the electron and the interaction potential  $V_{\rm int}(\rvec_{ij})=k_e e^2 / r_{ij}$. The direct interaction between nuclei is neglected in Eqn. \ref{eqn:electronhamiltonian}, and nuclei are static. Such an approximation is known as the Born-Oppenheimer approximation, and is common in band structure calculations. In the following, $a_{\rm nn}$ is the intersite spacing in the optical lattice, $a_{\rm nn,CM}$ is the interatomic spacing in the condensed matter system to be simulated, $M$ is the mass of the cold atoms in the simulator, and $m_{e}$ the mass of the electron. We use the intersite spacing instead of the lattice constant since there are 2 atoms per unit cell in graphene and BN, and blurring effects of the optical system will become relevant on the intersite distance before they are relevant on the scale of the lattice constant. This will also be the case for more complicated unit cells.

In a quantum simulator, interacting cold atoms, DRAs, ions or polar molecules with mass $M$ replace electrons, and an optical lattice replaces the periodic potential formed by nuclei, but the system naturally has a very similar Hamiltonian,
\begin{eqnarray}
& & -\frac{\hbar^2}{2M}\sum_{i}\nabla^2_{i}\psi^{(AT)} + \sum_{i\alpha}V^{(AT)}_{\rm ext}(\rvec_{i\alpha})\psi^{(AT)}\nonumber\\
 & &\;\;+ \sum_{i<j}V^{(AT)}_{\rm int}(\rvec_{ij})\psi^{(AT)}= E_{\rm AT} \psi^{(AT)}
\label{eqn:atomhamiltonian}
\end{eqnarray}
where $V^{(AT)}_{\rm ext}$ represents the external potential due to the optical lattice and $V^{(AT)}_{\rm int}$ is the interaction between the electron analogues. In an optical lattice system, the periodic potentials representing nuclei are genuinely static and the Born--Oppenheimer approximation is exact. This corresponds to neglect of phonons. Treating the effects of phonons with optical lattices is beyond the scope of this paper, and has been discussed elsewhere (See e.g. Ref. \onlinecite{hague2012a,hague2012b}).  For cold atoms, DRAs and cold ions, the interaction potential between the electron analogues may be significantly different in magnitude or form from the Coulomb interaction between electrons in a crystalline material.

The energy scales in the optical
system are necessarily lower, because the lattice has a larger size (with changes from
atomic scales $\sim 10^{-10}$m to optical length scales of between $\sim
10^{-7}$ and $10^{-6}$m)
and the electron analogues (the cold atoms, polar molecules or ions) are
heavier, which can be understood via a simple scaling argument. 

It can be helpful to rewrite the equation by introducing the dimensionless variable $r'$, related to $r$ via $r = a_{\rm nn,CM} r'$. In this form, 
\begin{displaymath}
-\frac{\hbar^2}{2m_e a_{\rm nn,CM}^2}\nabla'^2\psi'+V(a_{\rm nn,CM}r')\psi' = E_{\rm CM} \psi'
\end{displaymath}
hence, multiplying by a factor $m_e a_{\rm nn,CM}^2$ and rearranging, the equation reads:
\begin{equation}
-\frac{\hbar^2}{2}\nabla'^2\psi' = m_e a^2_{\rm nn,CM}\left(E-V(a_{\rm nn,CM}r')\right)\psi'
\end{equation}

The goal in the quantum simulator is to find a system with an identical second order differential equation.  The optical lattice has a different length scale, such that $r'_{AT}a_{\rm nn}=r_{AT}$. Performing the same analysis,
\begin{equation}
-\frac{\hbar^2}{2}\nabla'^2\psi'^{(AT)} = M a_{\rm nn}^2\left(E_{\rm AT}-V^{(AT)}(a_{\rm nn} r')\right)\psi'^{(AT)}
\end{equation}
so, in order for the prefactor on the right hand side of the equations to be the same, 
\begin{equation}
M a_{\rm nn}^2 E_{\rm AT} = m_e a_{\rm nn,CM}^2 E_{\rm CM}
\end{equation}
therefore the eigenvalues in the cold atom system are smaller by a factor of:
\begin{equation}
\frac{E_{\rm AT}}{E_{\rm CM}} = \frac{m_e a_{\rm nn,CM}^2}{M a_{\rm nn}^2}
\end{equation}
and a similar rescaling is needed for the potentials. Any rescaling of the wavefunctions cancels on both sides of the Schr\"odinger equation, and is not important to the discussion here.

An important consideration in the quantum simulator is the use of optical lattices to generate periodic
potentials for the cold atoms. The typical sinusoidal form of lattice
potential is very dissimilar from the attractive Coulomb potential generated by
nuclei. On the other hand, holograms \cite{bergamini2004a,nogrette2014a} and painted potentials
\cite{henderson2009a} offer the possibility to make custom optical
lattices.

A starting point for reproducing a Coulomb potential with optics is the convolution of a $-k_e e^2 Z / r_{\rm CM}$
potential with a Gaussian beam with waist, $w$, which has the form $\exp(-2r^2/w^2)$ \cite{self1983}. This convoluted potential has the
form, 
\begin{equation}
\tilde{V}_{\rm ext}(r_{\rm CM}) =-k_e e^2 Z \erf(r_{\rm CM}\sqrt{2}/w_{\rm CM})/r_{\rm CM}, 
\end{equation}
such that the $-k_e e^2 Z / r$ potential is recovered as $w\rightarrow 0$. This expression is written for condensed matter length scales. We require that $\tilde{V}/V_{\rm AT}=a^2_{\rm nn,CM}m_e/a_{\rm nn}^2 M$, $w_{\rm AT}=a_{\rm nn} w_{\rm CM}/a_{\rm nn,CM}$ and $r_{\rm AT}=a_{\rm nn} r_{\rm CM}/a_{\rm nn,CM}$ on rescaling for optical lattice length scales. Therefore the optical lattice potential should be,
\begin{eqnarray}
V^{\rm (AT)}_{ext}(r_{\rm AT}) & = & -k_e e^2 Z \frac{a^2_{\rm nn,CM}m_e}{a_{\rm nn}^2M}\frac{ \erf(r_{\rm AT}\sqrt{2}/w_{\rm AT})}{a_{\rm nn,CM} r_{\rm AT} / a_{\rm nn}}\\
& = & -k_e e^2 Z \frac{a_{\rm nn,CM}m_e}{a_{\rm nn} M}\frac{ \erf(r_{\rm AT}\sqrt{2}/w_{\rm AT})}{ r_{\rm AT} }
\label{eqn:opticallatpot}
\end{eqnarray}
The minimum value of $w=\lambda_{L}/2$, where $\lambda_{L}$ is
the wavelength of the light used to paint the optical lattice, i.e. the full beam width is $\sim \lambda_{L}$ (for recent experiments for addressing individual lattice sites, see e.g. Ref. \cite{parsons2015}). Thus, in any optical lattice system the sharp features of
the $-k_e e^2 Z / r$ potential are blurred.

\begin{figure}
\includegraphics[width=75mm]{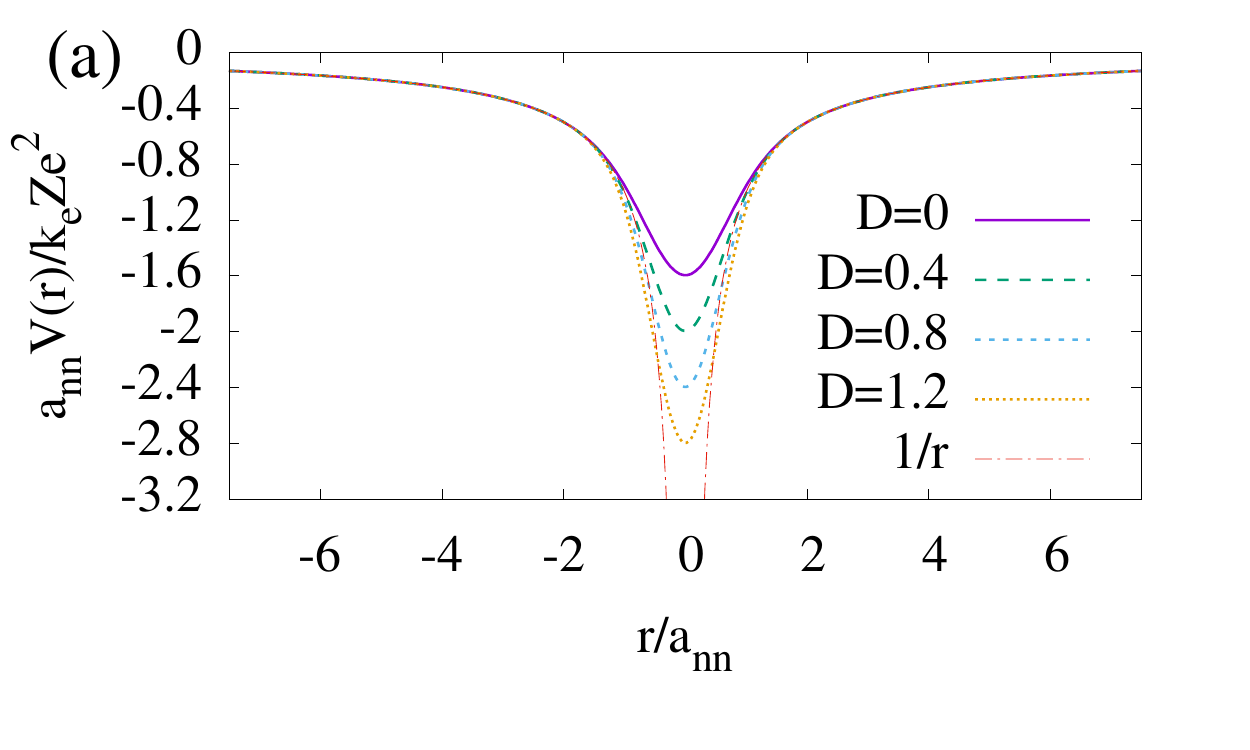}
\includegraphics[width=75mm]{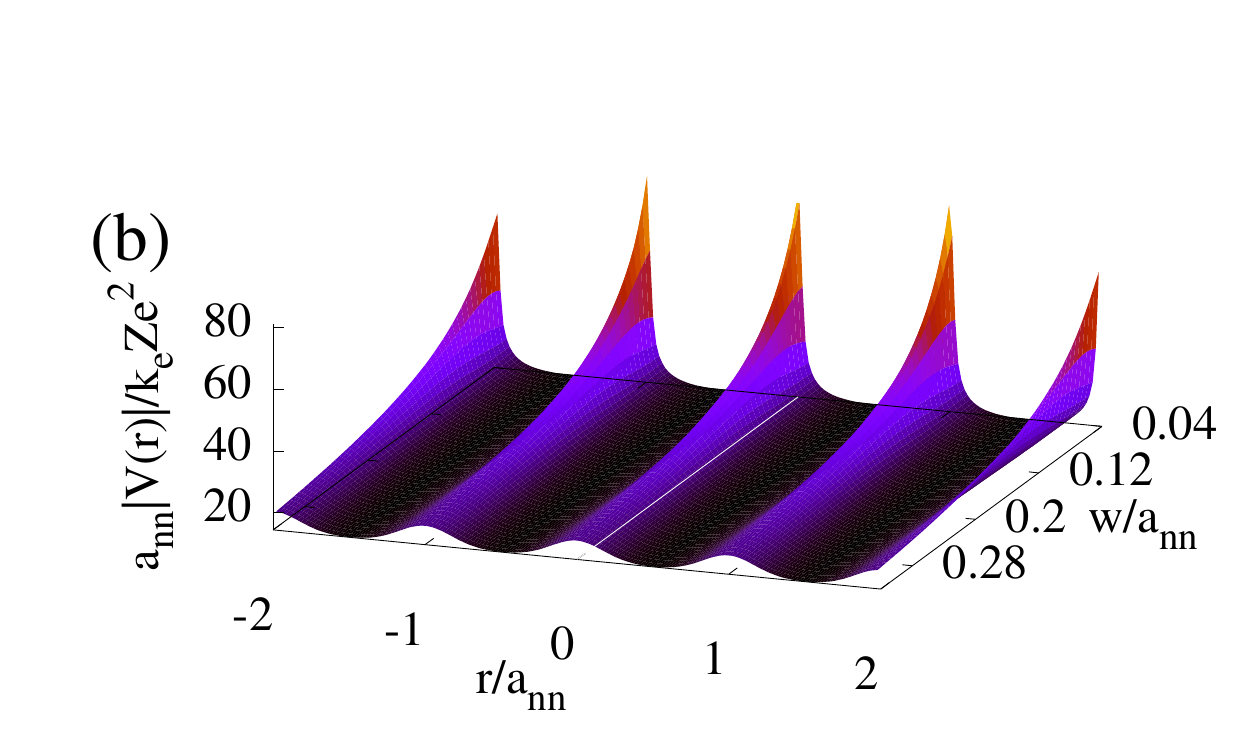}
\includegraphics[width=75mm]{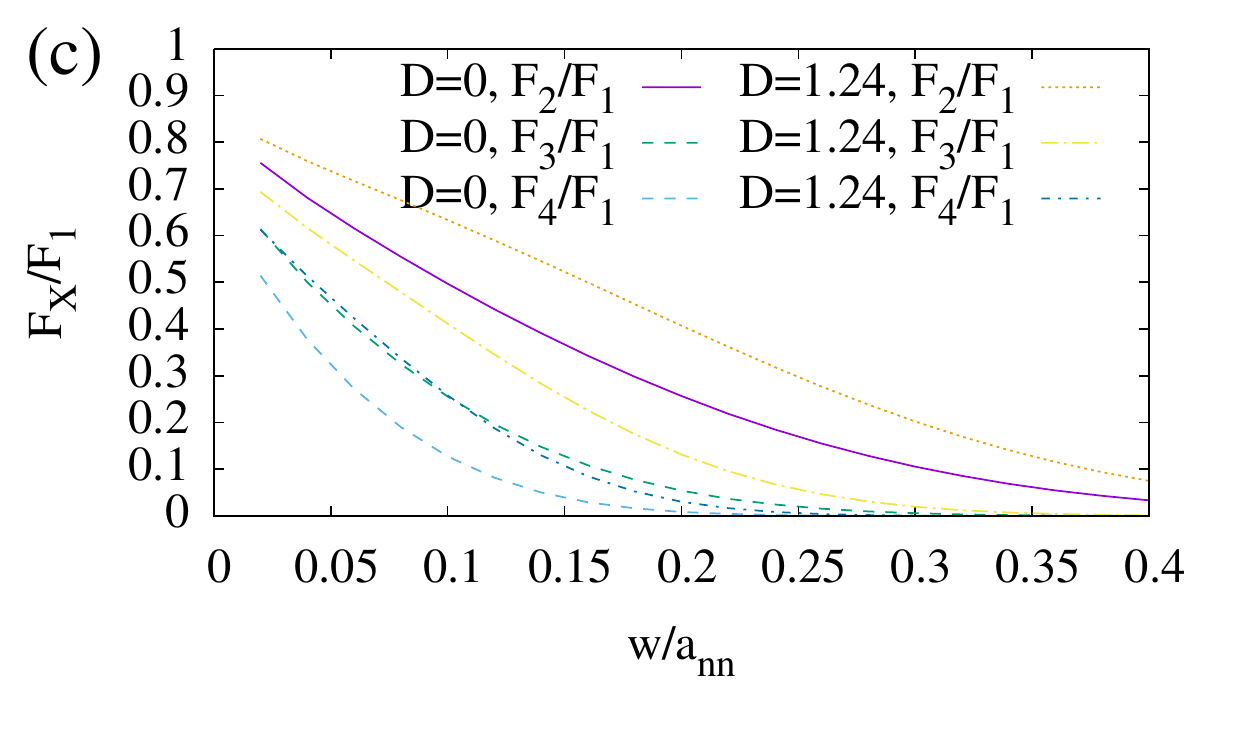}
\caption{(Color online) 
Panel (a) shows an example of a single Coulomb potential convoluted with
a Gaussian of waist $w=a_{\rm nn}$, and the same potential augmented with
an additional Gaussian beam at $r=0$, which makes the potential a better approximation to
the ideal $-Zk_{e} e^{2}/r$ form. Panel (b) shows a 1D periodic array of potentials,
augmented with a Gaussian at $r=0$ with $D=1.24$. Panel (c) shows the
Fourier cosine coefficients, $F_{X}$ of a 1D array of potentials with $D=0$ and $D=1.24$, to demonstrate how similar features in the optical lattices can be established using higher order spatial harmonics, where $X$ is the number of the coefficient. All quantities plotted are dimensionless.}
\label{fig:schematic}
\end{figure}

We also investigate the effects of an
additional augmentation where the beam intensity is increased at $r=0$ leading to the optical lattice potential,
\begin{equation}
V^{\rm (AT)}_{\rm ext}=-\frac{a_{\rm nn,CM}m_e}{a_{\rm nn} M}k_e e^2 Z \left[\frac{\erf(r\sqrt{2}/w)}{r}+\frac{D}{w}\exp\left(-\frac{2r^2}{w^2}\right)\right],
\label{eqn:extpot}
\end{equation}
which gives a slightly better approximation to the Coulomb potential if the dimensionless constant, $D$, is not too large. Note that there is an $a_{\rm nn,CM}m_e/a_{\rm nn}M$ factor associated with the second term, since $w$ also scales. A comparison between these potentials and the ideal Coulomb potential can be seen in Fig. \ref{fig:schematic}a. The curve marked $D=0$ shows the convolution between the Coulomb potential and Gaussian beam. As $D$ is increased up to around $D=1.2$, the range over which the exact Coulomb potential (labeled $1/r$) matches with Eq. \ref{eqn:extpot} increases significantly. This allows for far better agreement between lattice potentials without any requirement for smaller wavelength beams, and is a critical observation for the arguments presented in this paper. The nuclei in condensed matter systems are arranged in periodic arrays, and sums of these potentials can be seen in Fig. \ref{fig:schematic}b. These have a Coulomb form for small $w$, and eventually
gain a sinusoidal form as $w$ is increased. This can be more clearly seen in the Fourier coefficients of the array (Fig. \ref{fig:schematic}c), which could be used to identify simplified optical lattice forms for cases where $w$ is large enough that only a small number of spatial harmonics are needed to represent the external potential.

A similar argument requires the ratio of any energy scales associated with interactions between the electron analogues to be $m_{e}a_{\rm nn,CM}^2/Ma_{\rm nn}^2$. We will discuss interactions later in the paper.

As previously noted, there are two aspects to this paper. The first examines the effects of optical resolution on the band structure, making use of the local density approximation (LDA) of density
functional theory (DFT). Specifically, we use a modified version of the ELK
implementation \cite{elk} to examine the band
structures and particle density when the external potential has the form in Eqn. \ref{eqn:extpot}. ELK is an all-electron linear augmented plane
wave (LAPW) DFT code. For simplicity and necessity in the DFT calculations, we approximate the interaction between electron analogues with a $k_e e^2 / r$ form. This is accurate for cold ions. For cold Rydberg atoms or polar molecules, this gives us a rough method for analyzing the effects of the optical lattice resolution, but not the effect of non-Coulomb interactions on the band structure. Choosing a different form of interaction would involve the extremely complicated process of recomputing and parameterizing exchange-correlation potentials. The second aspect involving the calculation of dipole-dipole interaction strengths between dressed Rydberg atoms is carried out using van Vleck perturbation theory via a custom Mathematica script. In section \ref{sec:analogues} we discuss the ways in which differences in the interaction matrices between Coulomb and dipole-dipole interactions can be minimized. DFT calculations are used to examine the effects of changing interaction strength on the order of the difference between the Coulomb and Rydberg interaction matrices.

\section{Effect of optical lattice resolution}
\label{sec:resolution}


Calculations are made for low dimensional quantum
simulators for materials, since 2D lattices are the simplest to form using painted
potentials or holograms. The
two systems selected for study are a graphene simulator 
and a simulator for the closely related boron nitride,
since both materials have two dimensional
honeycomb lattices and weak electronic correlation. Numerical simulations of the quantum simulator are
made with increasing waist sizes $w$, until the band
structure near the Fermi surface fundamentally changes form. In the following, results labeled $w=0$ represent the full electron system calculated with the unmodified ELK code (i.e. the full condensed matter systems with an array of $-k_e e^2 Z/r$ nuclear potentials).

\begin{figure}
\includegraphics[width=0.445\textwidth]{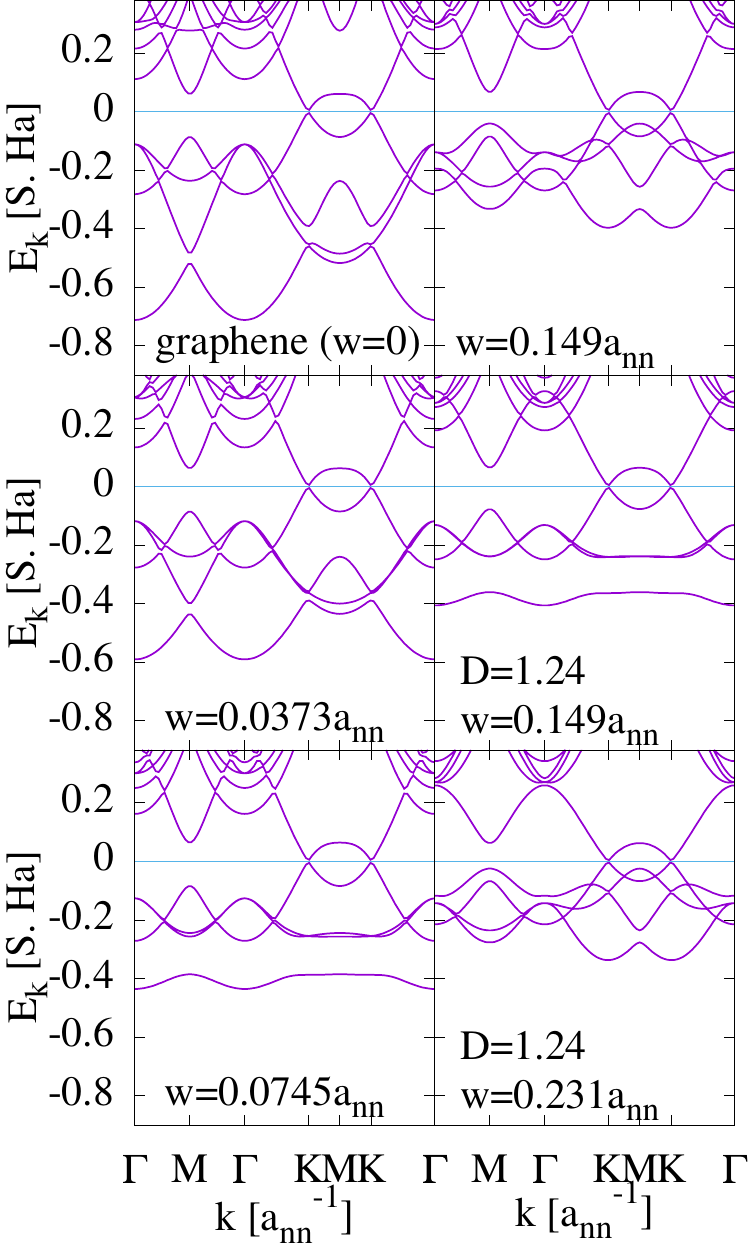}
\caption{(Color online) Electronic band structure of a quantum simulator for graphene with a variety of
  feature widths. For comparison, results from the unmodified ELK code are labeled $w=0$ (top left). The Gaussian waist modifies the band structure
  of the core states, and they increase in energy. Features close to
  the Fermi surface are well preserved until $w\sim 0.149a_{\rm nn}$, where a
  band starts increasing in energy at the M point, and heads towards
  the Fermi energy. Between $w=0.164a_{\rm nn}$ and $w=0.171a_{\rm nn}$, this band
  hits the Fermi surface, with the zero band gap semiconductor
  becoming a semi-metal. When $D$ is turned on a bigger waist size can be used. For $D=1.24$, the physics at the
  Fermi surface changes for $w\gtrsim 0.239a_{\rm nn}$. For the condensed matter system with $w=0$, energy has units of Hartrees (Ha) and for the optical lattice systems ($w\neq 0$) energy has units of $(m_{e}a_{\rm nn,CM}^2/Ma_{\rm nn}^2)$Hartree, which we will denote as scaled Hartrees (S. Ha). $1$ Hartree $\approx 27.21$eV. This convention is used throughout, to allow direct comparison between optical lattice and condensed matter systems.  $E_{k}$ is quasi-particle energy, $k$ is wavenumber and high symmetry points ($\Gamma$, $M$ and $K$) are defined in the usual way \cite{castroneto}.}
 \label{fig:graphene}
\end{figure}

Figure \ref{fig:graphene} shows the
predicted bandstructure of the graphene simulator for a set of different $w$. Results using the full electron local density approximation (LDA) of DFT calculated using the unmodified ELK code are shown towards the top left labeled as $w=0$. As the waist is
switched on, the band structure relating to the core states is
modified, and the core states increase in energy. As $w$ changes, the band structure maintains a similar structure close to the Fermi surface until $w\sim 0.149a_{\rm nn}$, where a
band below the Fermi surface exceeds the energy of the $\pi$ bands at
the M point, and heads towards the Fermi energy. Between $w=0.164a_{\rm nn}$
and $w=0.171a_{\rm nn}$, this band hits the Fermi surface, with the zero band
gap semiconductor becoming a semi-metal. The band structure near the Fermi surface maintains its form up to higher values of $w$ when the Gaussian correction to the optical lattice potential is introduced. For $D=1.24$, the physics at the
  Fermi surface changes between $w=0.231a_{\rm nn}$ and $w=0.239a_{\rm nn}$. It may be possible to use higher spatial harmonics to build up the optical lattice in place of painted potentials, reference to Fig. \ref{fig:schematic}(c) indicates that for $w\gtrsim 0.23a_{\rm nn}$ the second and third spatial harmonics of the Fourier series are
  important for obtaining the correct band structure, but that higher order spatial harmonics contribute only $\sim 1.5\%$ of the lattice shape and might be neglected. The relevance of this point to operating temperatures will be considered later in the paper. Since the core energies become higher than valence state energies in the limit that the lattice becomes sinusoidal, it is not possible to make a multiband quantum simulator simply by adding more atoms per site to the modified sinusoidal lattices described in e.g. Refs. \cite{tarruell2012,duca2015}. This demonstrates an advantage to using the more realistic optical lattice potentials proposed in this paper.

\begin{figure}
\includegraphics[width=0.45\textwidth]{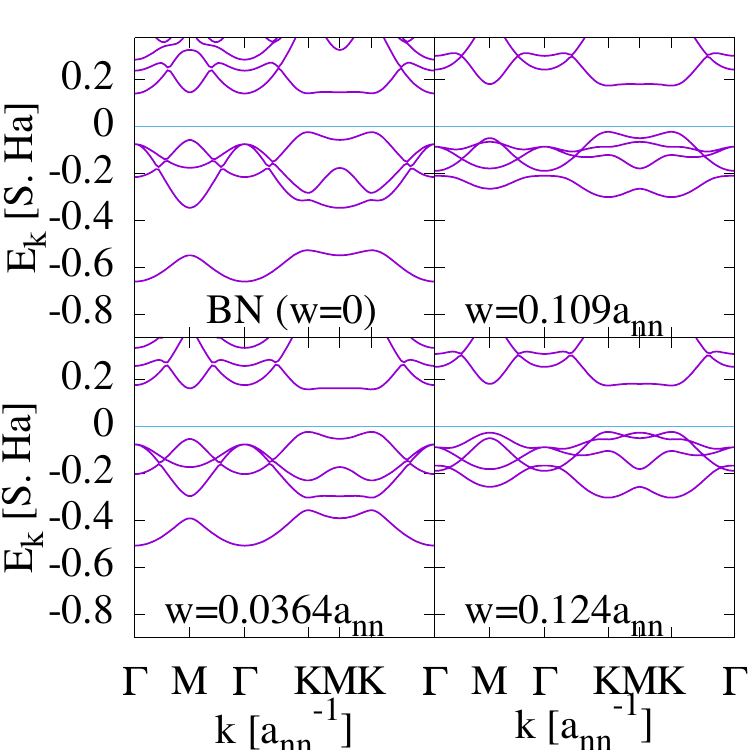}
\caption{(Color online) Electronic structure of a quantum simulator for BN with a variety of feature
  widths. Again, the results from the unmodified ELK code are shown, labeled as $w=0$. The bands start overlapping incorrectly at around
  $w/a_{\rm nn}=0.124$. The effects are particularly
  pronounced at the M point, and in the case of BN, the band structure
  goes from having a direct gap and the K point, to an indirect gap
  between M and K points. Quantities are defined as in Fig. \ref{fig:graphene}.}
\label{fig:boronnitride}
\end{figure}

As a zero bandgap semiconductor, graphene has states at the Fermi surface, but it is also instructive to examine a band insulator. Figure \ref{fig:boronnitride} shows the BN bandstructure for several waist sizes (again, the results from the unmodified ELK code are shown, labeled as $w=0$). The bands start overlapping incorrectly at around $w=0.124a_{\rm nn}$. The effects are particularly pronounced at the M
point, and in the case of BN, the band structure goes from having a
direct gap at the K point, to an indirect gap between M and K points.

\begin{figure}
\includegraphics[width=0.5\textwidth]{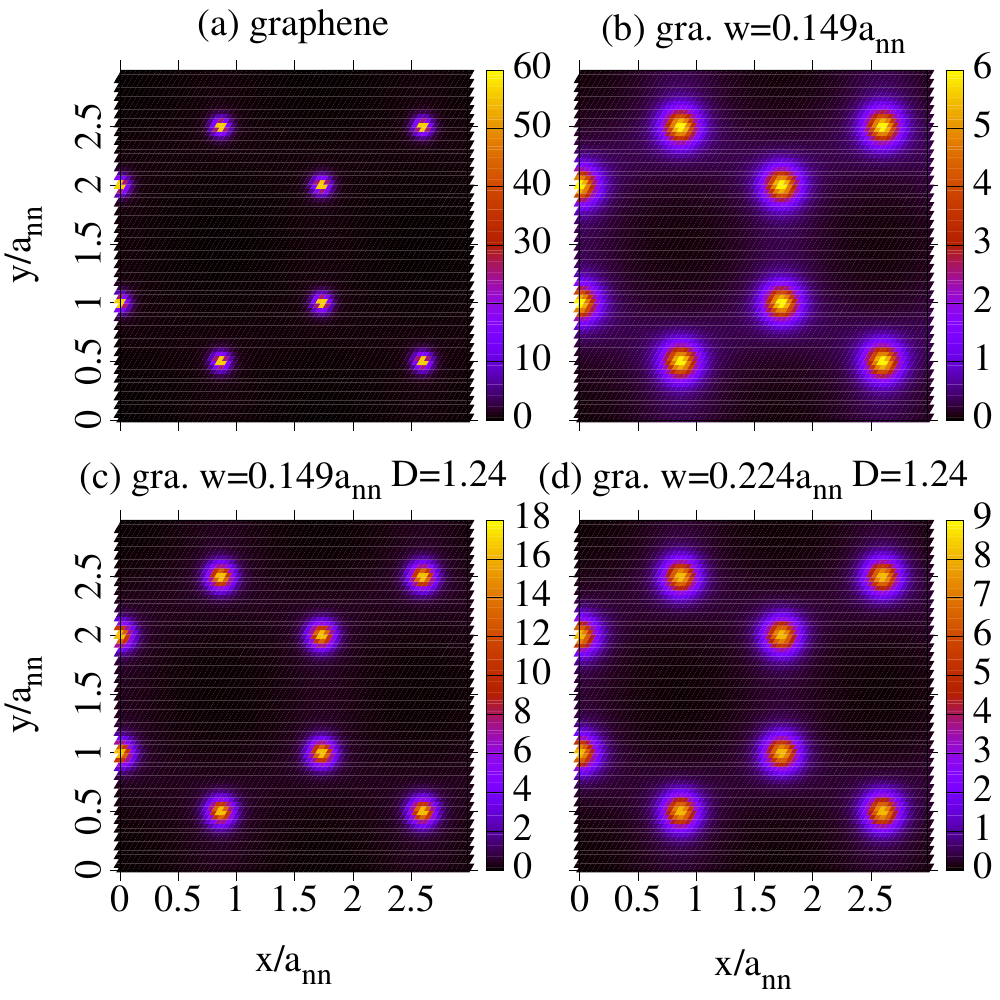}
\caption{(Color online) Charge
  density for graphene quantum simulator with (a) $w/a_{\rm nn}=0$ (b) $w/a_{\rm nn}=0.149$. Panels (c) and (d) show the effect of adding
  Gaussian corrections with $D=1.24$ to the lattice potential when $w/a_{\rm nn}=0.149$ and $w/a_{\rm nn}=0.224$. Inclusion of the Gaussian correction via $D$
  localizes atoms to lattice sites, leading to a better approximation
  to the condensed matter system.}
\label{fig:rhographene}
\end{figure}

\begin{figure}
\includegraphics[width=0.5\textwidth]{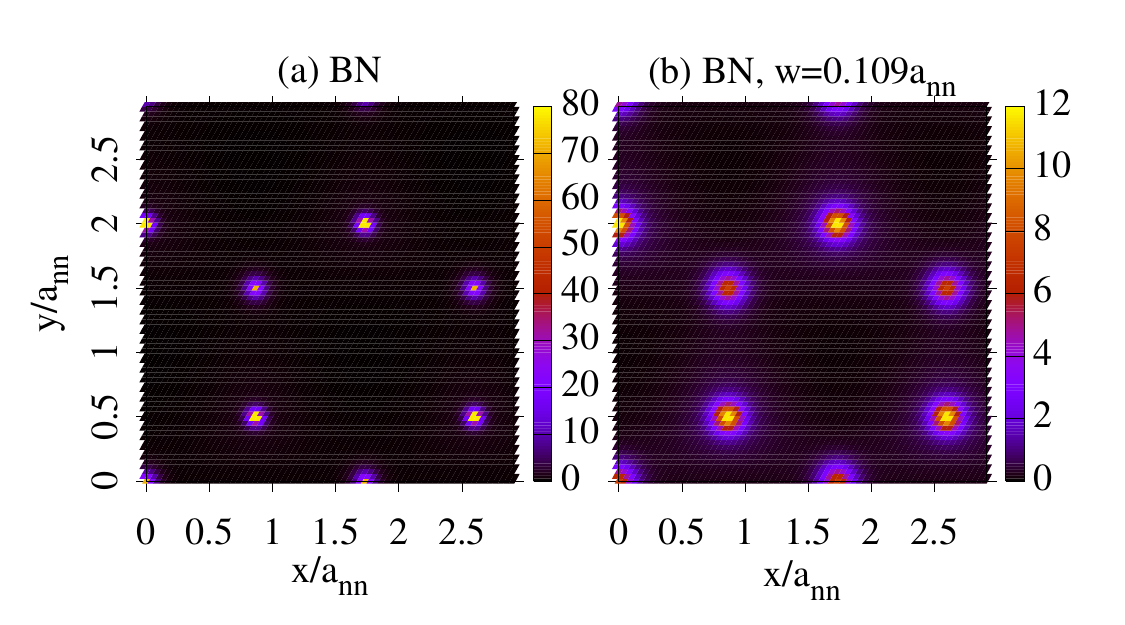}
\caption{(Color online) Charge
  density for BN quantum simulator with (a) $w/a_{\rm nn}=0$ (b) $w/a_{\rm nn}=0.109$.}
\label{fig:rhobn}
\end{figure}

Figure \ref{fig:rhographene} shows the electron analogue density for
extreme values of $w$ for the graphene quantum simulators. The atom
(electron) density is much more spread out for larger waist
sizes. When the additional Gaussian beam at $r=0$ is added, the particle
density becomes more localized, but the band structure at the Fermi
surface (Figs. \ref{fig:graphene} and \ref{fig:boronnitride}) breaks
down when the particle density around the nuclei obtain a similar
size. This can also be seen for the BN analogue in Fig. \ref{fig:rhobn}. It is clear that having an appropriately localized electron
analogue density is key to obtaining the correct band structure near the Fermi surface,
however, it is also important that the lattice potential is not
significantly deeper than $-k_e a_{\rm nn,CM} m_e e^2 Z / a_{\rm nn} M r$ (which would also localize
the electron analogues), as this can also lead to significant changes in the band structure.

\section{Interactions between electron analogues}
\label{sec:analogues}

The analysis in the previous section has implications for the smallest size of the intersite distance, and due to the energy scaling associated with change of length scales will affect the size of interaction that is required between electron analogues. The smaller the lattice constant, the larger the energy scaling factor $m_e a_{\rm nn,CM}^{2}/Ma_{\rm nn}^2$ and thus the higher the energy scales (and operating temperature). However, the required interaction strength also increases as length scales decrease, which could increase decoherence (see Sec. \ref{sec:coherence}). This indicates a trade-off between achievable interaction strengths and operating temperatures. Before we continue, we estimate the shortest possible distance between lattice sites.

As we have previously stated, $w=\lambda_{L}/2$. According to Sec. \ref{sec:resolution}, the smallest distance between lattice sites that can be used to reproduce the physics of the Fermi surface is found from $w=0.231a_{\rm nn,best}$ for a graphene simulator when $D=1.24$, corresponding to $a_{\rm nn,best}=\lambda_{L}/(2 \times 0.231) = 2.16\lambda_{L}$. For a blue-green laser, $\lambda_{L}=532$nm corresponding to $a_{\rm nn,best}=1.15\mu\mathrm{m}$. We will use this value of $\lambda_{L}$ throughout.

Examination of the spatial harmonics of the Fourier series indicates that it might be possible to use lattices with even smaller constants. While we are unable to directly analyze sinusoidal lattices using the ELK code, the Fourier analysis in Fig. \ref{fig:schematic}(c) indicates that only the fundamental mode plus two spatial harmonics is needed to reproduce the optical lattice for $w\sim 0.23a_{\rm nn}$ to around 1.5\% accuracy. 

It would be exceptionally challenging to build the Fourier series using a superposition of standing waves formed with lasers with different wavelengths. However, it might be possible to use holographic potentials to introduce multiple scales. For the three spatial harmonics needed to reproduce the lattice, the fundamental mode would be chosen to have lattice constant $3\lambda_{L}/2$, the second harmonic $\lambda_{L}$ and the third harmonic to have the smallest possible spacing of $\lambda_{L}/2$.\footnote{There are good reasons to believe that the smallest possible spacing between intensity maxima for simple sinusoidal optical lattices is $\lambda_{L}/2$, regardless of the method used to form them. For example, optical lattices built from standing waves have lattice constant $\lambda_{L}/2$, and in the limit that Gaussian beams are painted very close together, it is straightforward to check that the smallest lattice constant of the resulting sinusoidal lattice is $\sim w=\lambda_{L}/2$.}

We therefore take a lattice constant of $\bar{a}_{\rm nn,best}=3\lambda_{L}/2$ as an absolute limit for the optical lattice built from spatial harmonics (which is slightly smaller than the painted potential system). This corresponds to $\bar{a}_{\rm nn,best}=798$nm using a blue-green laser \footnote{We also note that depending on whether the lasers are red or blue detuned relative to the energy spacing of particular atoms, the lattice will be attractive (red detuned) or repulsive (blue detuned). In the latter case, holes will be made in the lattice, rather than depressions, but the theoretical analysis still applies.}. We present results for both lattice sizes, and the conclusions of this paper do not rely on this smaller lattice constant. Rather, the implementation of smaller lattices using Fourier components may provide an interesting route for improving experimental implementations.

The interaction between the electron analogues must be tuneable and close to the values required by the scaling argument of Sec. \ref{sec:modelmethod}. Here, we mainly focus on dressed Rydberg atoms and briefly discuss the relative advantages and disadvantages of systems of polar molecules, cold atoms interacting via Feshbach resonances and cold ions. For the purpose of calculating the scaling factors, the interatomic spacing in graphene is $a_{\rm nn,CM}=1.42$\AA.

\begin{figure}
\includegraphics[width=0.45\textwidth]{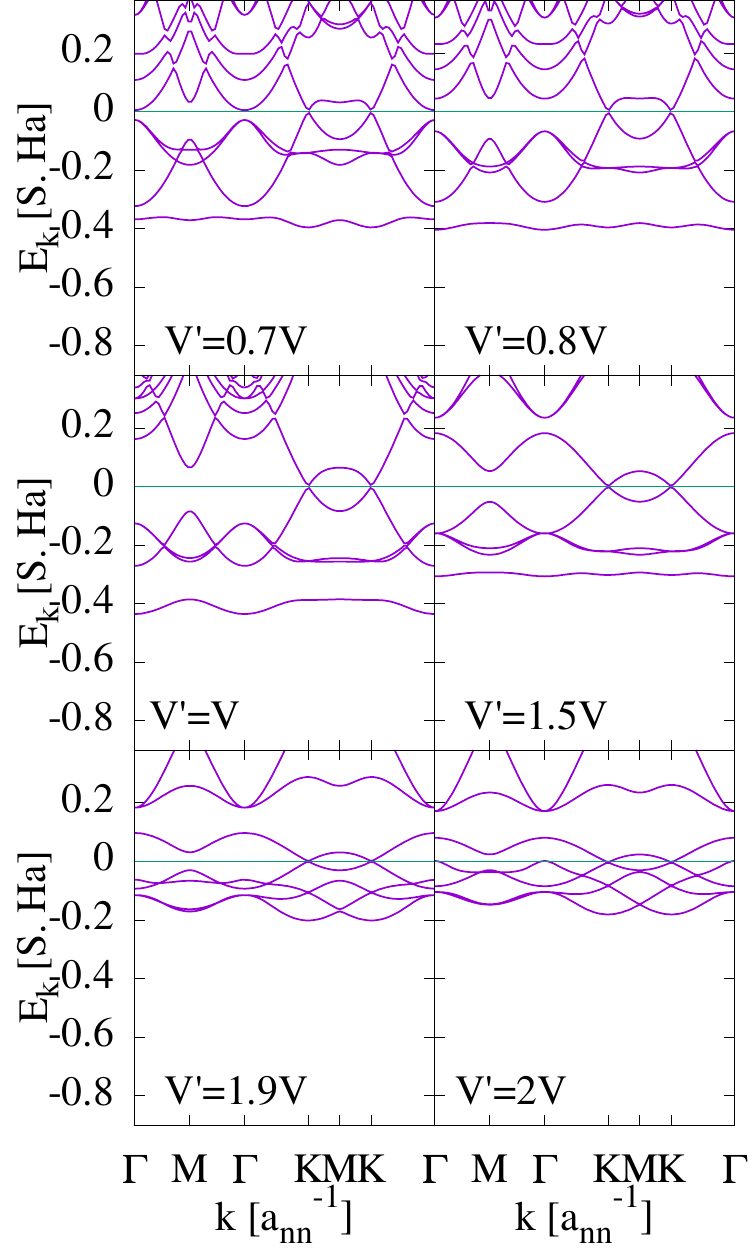}
\caption{Effect of scaling the interaction on the band structure of simulated graphene. Here, $w=0.0745a_{\rm nn}$. The K points are well defined between $V'=0.8V$ and $V'=1.9V$, a scaling factor a little over 2, with the separate 1s core state separated from the $\pi$ and sp$^{2}$ bands until $V'=1.5V$. The main effect of increased interaction is an increase in the energy of the core band, and a decrease in bandwidth size indicating greater localization of the free particles. Quantities are defined as in Fig. \ref{fig:graphene}.}
\label{fig:interactionscale}
\end{figure}

The use of non-Coulomb interactions between electron analogues depends critically on the sensitivity of the band structure to the interaction strength. We can estimate the modification of the band structure in response to changes in the interaction potential using DFT, which is shown in Fig. \ref{fig:interactionscale}. A linear dispersion around the K points can be found for interactions between $V'=0.8V$ and $V'=1.9V$, a scaling factor a little over 2. Here, $V$ is the unscaled interaction strength. One of the main effects of increased interaction is an increase in the energy of the core band, and a decrease in the widths of the bands, consistent with greater localization of the electron analogues. This indicates that the qualitative features of the band structure could emerge in a multi-band quantum simulator even if the interaction strengths are only approximately similar.

\subsection{Dressed Rydberg atoms}
\label{sec:rydberg}

In this section, we compare the size of the interaction matrices between DRAs with those for the Coulomb interaction. Rydberg atoms interact by long range van der Waals forces. By tuning a laser close to the ground state to Rydberg state transition, we can make a superposition of the ground and Rydberg states, where the amplitude of the Rydberg state, $\alpha$, is kept small. In this case, the mostly ground state atoms can interact via the Rydberg states. If the Rabi frequency of the atom-laser interaction is $\Omega_{\mathrm{2ph}}$ and the detuning of the laser from the ground-Rydberg state transition is $\Delta_{\mathrm{2ph}}$, then $\alpha=\Omega_{\mathrm{2ph}}/2\Delta_{\mathrm{2ph}}$.  This means the interaction is laser controlled, and we benefit by suppressing the effect of the finite Rydberg state lifetime,  since the lifetime of the superposition state is longer by a factor of $1/\alpha^2$, i.e. the probability of finding the atom in the Rydberg state is a factor $\alpha^2$ smaller. In order to assess the interaction between DRAs, we have chosen $^{43}$Ca and $^{87}$Sr which have $4s^2$ and $5s^2$ ground states respectively, because they are fermionic, have readily available data, have reasonable $C$ coefficients (which we compute following Ref. \cite{vaillant2012}), can be excited to $4sn_{\rm Ryd}s$ or $5sn_{\rm Ryd}s$ Rydberg states with nearly spherically symmetric interactions and because there is exists a transition with extremely long lifetime which greatly suppresses heating in the Rydberg dressing scheme (see Sec. \ref{sec:coherence}). We note that lighter alkaline earth metals are likely to have similar transitions and narrow linewidths and might have other advantages (e.g. higher operating temperature). 

\begin{figure}
\includegraphics[width=0.45\textwidth]{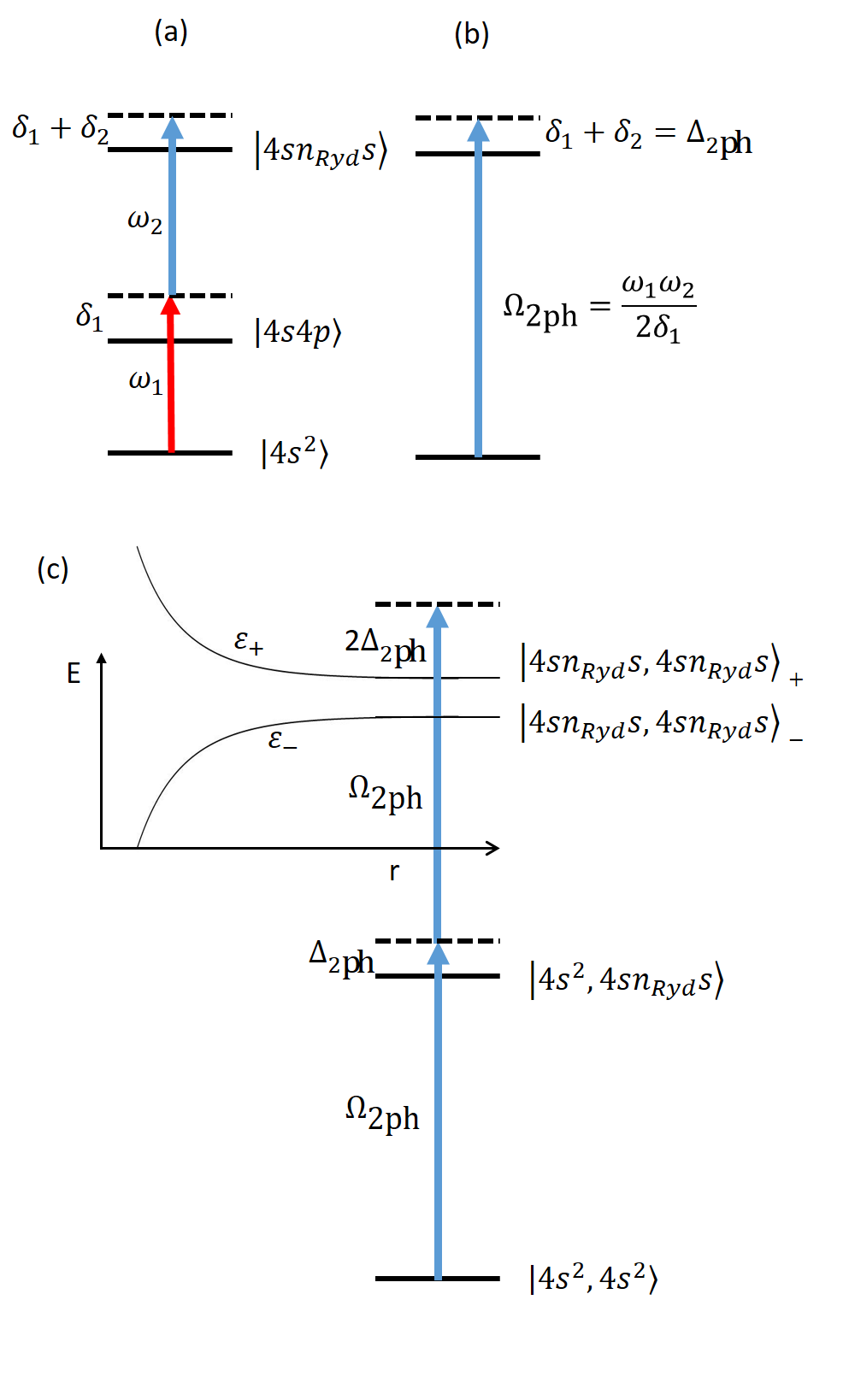}
\caption{Scheme used to dress the atoms with lasers. (a) shows the relevant transitions for driving a single $^{43}$Ca or $^{87}$Sr atom via a two-photon transition, with Rabi frequencies $\omega_i$ and detunings $\delta_i$, where i=1,2 labels the transitions. (b) the two photon transition is equivalent to a single transition driven with Rabi frequency $\Omega_{\textrm{2ph}}$ and detuning $\Delta_{\textrm{2ph}}$. (c) For two atoms, the excitation to the 2 atom states has an energy spectrum modified by the Rydberg-Rydberg interaction. The two-photon transition dresses the higher energy 2 Rydberg-atom state, being far detuned from the lower energy 2 Rydberg atom state.}
\label{fig:dressingscheme}
\end{figure}

For almost spherically symmetric interactions we wish to excite the $4sn_{\rm Ryd}s$ or $5sn_{\rm Ryd}s$ Rydberg states respectively, where $n_{\rm Ryd}$ is the principal quantum number of the Rydberg state, and this means the laser dressing must be via two photons. The first photon is from a laser tuned close to the narrow linewidth $^1S_0\rightarrow ^3P_1$ ``intercombination line'' with Rabi frequency $\omega_1$ and detuning $\delta_1$ (for $^{43}$Ca the linewidth is $2\pi \times 6.5 \; \mathrm{kHz}$ and wavelength 657 nm; for $^{87}$Sr the linewidth is $2\pi \times 7.5 \; \mathrm{kHz}$ and wavelength 689 nm). The second photon is from a laser tuned close to the $^{43}$Ca $4s4p \rightarrow 4sn_{\rm Ryd}s$ transition (or the $^{87}$Sr $5s5p\rightarrow 5sn_{\rm Ryd}s$ transition), with Rabi frequency $\omega_2$ and detuning $\delta_2$. The effective ``two photon'' Rabi frequency is then $\Omega_{\mathrm{2ph}}=\omega_1 \omega_2/2\delta_1$, and the two-photon detuning $\Delta_{\mathrm{2ph}} = \delta_1+\delta_2$ \cite{henkel2010}. A schematic of the dressing scheme for $^{43}$Ca is shown in Fig. \ref{fig:dressingscheme}. The $^{87}$Sr dressing scheme is similar. The interactions between atoms are fully determined from $\Delta_{\rm 2ph}$ and $\Omega_{\rm 2ph}$, however the values of $\delta_i$ and $\omega_i$ can affect heating (see Sec. \ref{sec:coherence}).  The interactions between dressed Rydberg atoms are highly tuneable and with the right dressing scheme, almost spherically symmetric interactions between the dipole moments of the atoms can be induced \footnote{We note the nearby and exceptionally narrow clock transition in Sr (I$^1S_0\rightarrow ^3P_0$ ``intercombination line'') which is of high current interest for optical lattice clocks, however it is so narrow that unreasonably high power lasers would be needed to implement our scheme.}.

For simplicity, we consider the interactions of two atoms in the presence of the dressing light. The dressed Rydberg atom interaction potential can be calculated using e.g. van Vleck perturbation theory \cite{vanvleck1929}, which gives,
\begin{equation}
V^{\mathrm{Tot}}_{\pm} = \frac{(\epsilon_{\pm}-2\Delta_{\mathrm{2ph}})\Delta_{\mathrm{2ph}} \Omega_{\mathrm{2ph}}^2 + \Omega_{\mathrm{2ph}}^4}{\Delta_{\mathrm{2ph}}(4\Delta_{\mathrm{2ph}}^2-\Omega_{\mathrm{2ph}}^2-2\epsilon_{\pm}\Delta_{\mathrm{2ph}})}.
\end{equation}
where $\epsilon_{\pm}$ represent the bare Rydberg-Rydberg interaction (see below). In the limit where $\epsilon_{\pm}\rightarrow 0$, $V^{\mathrm{Tot}}_{\pm}\rightarrow V_L=(\Omega_{\rm 2ph} ^4-2 \Delta_{\rm 2ph} ^2 \Omega_{\rm 2ph} ^2)/(4 \Delta_{\rm 2ph} ^3-\Delta_{\rm 2ph}  \Omega_{\rm 2ph} ^2)$, which is due only to the laser dressing, but not the interactions of the atoms. We subtract $V_L$ from $V^{\mathrm{Tot}}_{\pm}$ to obtain the effective interaction potential:
\begin{equation}
V_{\pm}=\frac{\epsilon_{\pm} \Omega_{\mathrm{2ph}} ^4}{\left(\Omega_{\mathrm{2ph}} ^2-4 \Delta_{\mathrm{2ph}} ^2\right) \left(2 \epsilon_{\pm} \Delta_{\mathrm{2ph}} +\Omega_{\mathrm{2ph}} ^2-4 \Delta_{\mathrm{2ph}} ^2\right)}.
\end{equation}
When the separation between the two atoms is small, the shift of the Rydberg levels by the interaction inhibits the excitation of the Rydberg levels by the laser, and so the interaction saturates at $-\Omega_{\mathrm{2ph}} ^4/(8 \Delta_{\mathrm{2ph}} ^3-2 \Delta_{\mathrm{2ph}}  \Omega_{\mathrm{2ph}} ^2)$, which is $\approx\Omega_{\mathrm{2ph}}\alpha^3$ when $\Delta_{\mathrm{2ph}}\gg\Omega_{\mathrm{2ph}}$. 
To calculate $\epsilon_{\pm}$, we must diagonalize the Hamiltonian for the two-atom states, which takes the form of a $2\times2$ matrix in the basis $|n_{\rm Ryd}s,n_{\rm Ryd}s\rangle$, $|n_{\rm Ryd}^{\prime}p, n_{\rm Ryd}^{\prime\prime}p\rangle$:
\begin{equation}
\left(
\begin{array}{cc}
0 & C/r^3 \\ C/r^3 & \delta_{\rm For}
\end{array}
\right)
\end{equation}
where $\delta_{\rm For} = (E_{n_{\rm Ryd}^{\prime}p}+E_{n_{\rm Ryd}^{\prime\prime}p})-2E_{n_{\rm Ryd}s}$ is the Forster defect, and $E_{n_{\rm Ryd}l}$ is the energy of $n_{\rm Ryd}l$ state of the atom. In the dressed Rydberg scheme, the laser driven transition is always much closer to the state with the eigen-value $\epsilon_{+}>0$, which ensures the interactions are repulsive. Therefore, in the following, we will always be discussing $V_+$ when we refer to the interaction potential.
The $C$ coefficient is determined by the Rydberg states involved in the scattering, following Ref. \cite{vaillant2012}. We need to choose an interaction that will mimic the scaled interaction matrix strength, which depends on the optical lattice constant and the mass of the atom used in the simulator. At the short distances over which the interaction potential is significant (up to about 2 microns), the potential can be simplified because $\epsilon_+\approx C/r^3$ when r is small. It follows that 
\begin{equation}
V_+=\frac{\alpha^4}{(\alpha^2-1)^2}\frac{C}{r^3+C\alpha/\Omega_{\mathrm{2ph}}(\alpha^2-1)},
\end{equation}
so $V_+$ has the form $V_+ = V_0/(r_c^3+r^3)$, with $r_c = \left(C\alpha/\Omega_{\mathrm{2ph}}(\alpha^2-1)\right)^{1/3}$. N.B. Both $\alpha^2-1$ and $\alpha$ are negative in the following, so $r_{c}$ is positive.

\begin{table*}
\caption{Parameters for proposed experimental DRA quantum simulator. The upper part of the table summarize the parameters of the quantum simulator for the specified atom, and the properties of the transitions that should be driven. The last 2 column are for $^{25}$Mg, for which we do not have parameters for the Rydberg levels, so we estimate using $C$ and $\delta_{\rm For}$ for Ca (see text). We select the same Rabi frequencies for both transitions required for the two-photon scheme, so  that $\omega_2 = \omega_1$, and the detuning for the upper transition satisfies $\delta_2 = \Delta_{2ph} - \delta_1 $. The lower rows of the table relate to decoherence effects and are discussed in Sec. \ref{sec:coherence}. We also show the laser power needed to drive the lower $^1S_0 \to ^3P_1$ ``intercombination'' line transition, assuming that the laser is focused into either a $50\times50\;\mu$m area (for $^{43}$Ca and $^{87}$Sr) or $25\times25\;\mu$m for ($^{25}$Mg). The lowest row gives an estimate of the number of hops per simulation, which is calculated from either the DRA lifetime or the heating (whichever is smaller).}
\label{tab:draparams}
\begin {tabular} {c c c c c c c}
\hline
Species & \multicolumn{2}{c}{$^{43}$Ca} & \multicolumn{2}{c}{$^{87}$Sr} & \multicolumn{2}{c}{$^{25}$Mg} \\
\hline
Transition & \multicolumn{2}{c}{$4s^2\rightarrow 4s4p \rightarrow 4sn_{\rm Ryd}s$} & \multicolumn{2}{c}{$5s^2\rightarrow 5s5p \rightarrow 5sn_{\rm Ryd}s$} & \multicolumn{2}{c}{$3s^2\rightarrow 3s3p \rightarrow 3sn_{\rm Ryd}s$}\\
\hline
$n_{\rm Ryd}$ & 32 & 38 & 32 & 38 & 32 & 38 \\
$C$ (MHz $\mu$m$^3$) & 24.7 & 50.1 & 29.3 & 63.3  & \multicolumn{2}{c}{Estimate from Ca} \\
$\delta_{\textrm{For}} \; \textrm{(GHz)}$ & -16.700 & -8.880 & -10.900 & -5.440  &  \multicolumn{2}{c}{Estimate from Ca} \\
$\lambda_{^3P_1} \; \textrm{(nm)}$ & 657 & 657 & 689 & 689 & 457 & 457 \\
$\Gamma_{^3P_1} \; \textrm{(Hz)} $ & 6000 & 6000 & 7460 & 7460 & 31 & 31 \\
$I_{\textrm{Sat} \; ^3P_1} \; \mu \textrm{W/cm}^2 $ & 2.770 & 2.770 & 2.980 & 2.980 & 0.042 & 0.042 \\
\hline
Inter-site distance $a_{\textrm{nn}}$ (nm) & 798 &  1150 & 798 & 1150 & 798 & 1150\\
$\Omega_{2\textrm{ph}}$ (MHz) & 6.17 & 2.55 & 6.51 & 2.55 & 7.07 & 2.92 \\
$\Delta_{2\textrm{ph}} $ (MHz) & -34.29 & -14.17 & -46.48 & -17.96 & -35.35 & -14.60 \\
$\alpha $ & -0.09 & -0.09 & -0.07 & -0.071 & -0.1 & -0.1\\
 $1/\alpha^2$ & 123.5 & 123.5 & 204.1 & 198.4 & 100 & 100 \\
Effective Lifetime Ryd. $\alpha^{-2}\tau_{\textrm{Ryd}}$ (ms) & 1.57 & 2.63 & 2.59 & 4.22 & 1.27 & 2.13\\
\hline
detuning $ns^2 \to  nsnp$: $\delta_1$ (GHz) & 9.19 & 9.19 & 9.19 & 9.19 & 0.50 & 0.50 \\
Rabi freq. $ns^2 \to nsnp$: $\omega_1$ (MHz) & 337 & 217 & 346 & 217 & 84 & 54\\
Laser power (mW) & 436 & 180 & 320 & 126 & 3900 & 1610\\
DRA heating rate $\dot{E}_{\rm Ryd}$ (nK/ms) & 13.0 & 5.4 & 7.6 & 3.0 & 5.0 & 2.1\\
\hline
$\lambda_{\rm eg}$ (nm) & 423 & 423 & 460 & 460 & 285 & 285 \\
$\Gamma_{\rm eg}$ (GHz) &  0.2 & 0.2 & 0.217 & 0.217 &  0.5 & 0.5 \\ 
$\lambda_{L}$ (nm) & 532 & 532 & 532 & 532 & 532 & 532 \\
Lat. heating rate, $\dot{E}_{L}$ (nK/ms) & 0.56 & 0.39 & 0.21 & 0.14 & 1.13 & 0.78 \\
\hline
Total heating rate, $\dot{E}$ (nK/ms) & 13.56 & 5.79 & 7.81 & 3.14 & 6.13 & 2.88 \\
Hopping time $\tau $ (ms) & 0.20 & 0.4 & 0.41 & 0.85 & 0.12 & 0.24\\
$T_{10\%}$ (nK) & 8.43 & 4.05 & 4.17 & 2.00 & 14.4 & 7.2\\
\hline
hops/sim. $\approx T_{10\%}/\dot{E} \tau $ or $\tau_{\rm Ryd}/\tau\alpha^2$ & 3.11 & 1.76 & 1.29 & 0.75 & 10.58 & 8.875 \\
\hline
\end {tabular}
\end{table*}

We proceed to estimate the strengths of the interaction matrices for such an interaction, which have the form,
\begin{equation}
U = \int \dr^{3}\rvec \dr^{3}\rvec'  V(\rvec-\rvec') N(\rvec) N'(\rvec')
\end{equation}
We assume that the electron (electron analogue) density is step like, $N(\rvec)=N \,(r<b)$ and $N(\rvec)=0$ otherwise, approximating the form of a Wannier function. Therefore, $N=3/4\pi b^3$. For in-band interactions (which are largest), $N=N'$. Interband interactions are smaller. Matching of the interactions on the scale of lattice spacing requires that $U/U_{coul} = m_{e}a_{\rm nn,CM}^2/Ma_{\rm nn}^2$. The experimental aim would be to get a match within a range of a few 10s of percent, so that changes in interaction do not strongly change the band structure (see also Fig. \ref{fig:interactionscale}).

\begin{figure}
\includegraphics[width=0.48\textwidth]{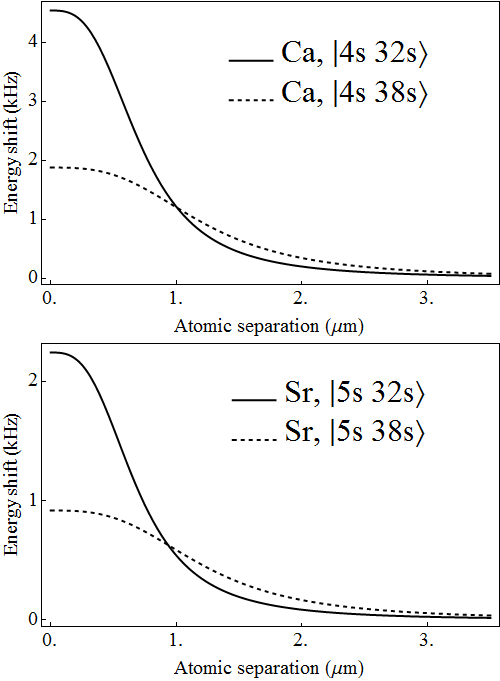}
\caption{Interactions between dressed Rydberg states of pairs of $^{43}$Ca atoms, and pairs of $^{87}$Sr atoms. Coefficients are as in Table \ref{tab:draparams}.}
\label{fig:interaction}
\end{figure}

\begin{figure}
\includegraphics[width=0.47\textwidth]{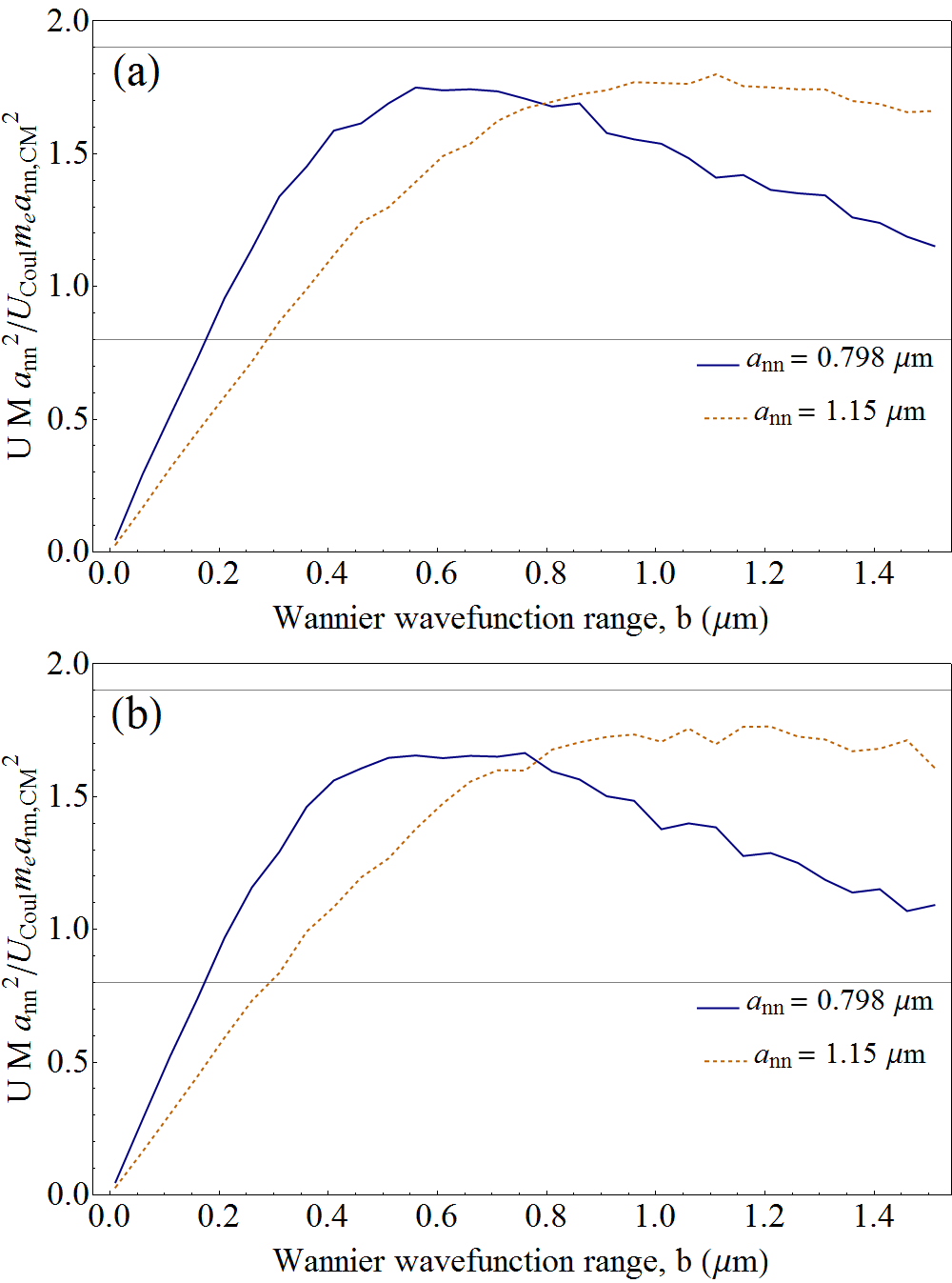}
\caption{Ratio of the interaction matrix elements of Rydberg atoms to Coulomb interactions calculated using Monte Carlo integration for (a) $^{43}$Ca and (b) $^{87}$Sr. The horizontal lines show the maximum and minimum bounds on the interaction strength determined using DFT.}
\label{fig:integralplot}
\end{figure}

By modifying the principal quantum number of the Rydberg state, it is possible to modify the range of distances over which the interaction matrices for both Coulomb and dipole-dipole interactions have roughly the same magnitude. We change the strength of the interaction by varying $\Delta_{\rm 2ph}$, $n_{\rm Ryd}$ and $\Omega_{\rm 2ph}$. A summary of the values used can be found at the top of Table \ref{tab:draparams}. The form of the interactions for the different parameter sets is shown in Fig. \ref{fig:interaction}.  Numerical evaluations of the interaction matrix elements using Mathematica's Monte Carlo integration routines for the four sets of parameters can be seen in Fig. \ref{fig:integralplot}. The figures also show the rough bounds on the interaction matrices determined by the DFT calculations in Fig. \ref{fig:interactionscale}.

It can be seen that while the interactions do not have a Coulomb form, the interaction matrices have magnitudes within the bounds indicated by the DFT calculations over a range of Wannier-like function sizes, from radii slightly larger than those shown in the figure, down to around 25\% of the largest sizes shown. The reason for the region of approximate agreement is that the plateau in the interaction at small $r$ means that Coulomb and dipole-dipole interactions have a region where they are tangential. An experimental protocol would aim to match  interaction matrices for Wannier-like functions on the size of an interatomic spacing (i.e. those most associated with the Fermi surface) by tuning the dipole-dipole interaction through the Rabi frequency, detuning parameter, and principal quantum number of the Rydberg state to maximize the length scales over which the interaction matrices are of similar magnitude. This is likely to be problem specific.

We can speculate about how this might affect different states by using hydrogenic wavefunctions as an estimate. The radius of such wavefunctions scales roughly as $n_{\rm Wann}^2$. Hence $n_{\rm Wann}=1$ states are approximately $1/4$ of the size of $n_{\rm Wann}=2$ states, where $n_{\rm Wann}$ is the principal quantum number representative of the Wannier state (not to be confused with the principal quantum number of the Rydberg state). The results indicate that the interaction matrices of materials containing elements in the first 2 rows of the periodic table could be approximately represented using the types of DRAs we have discussed here, although there may be difficulties with quantum simulation of materials containing elements in the 3rd and higher rows. This difficulty is partly related to heating, which we discuss later, and might be avoided by using different Rydberg dressing schemes to the ones discussed here. The interactions associated with electron analogues in the smaller core states would be underestimated, and probably the energies of these states would be lowered. The Rydberg dressing scheme would need careful tuning for each problem to be investigated, so results from such a quantum simulator may be able to give useful insight into specific physical processes in materials, but would not be predictive.

\subsection{Polar molecules}

For completeness, we mention an intriguing possibility to use polar molecules. The
recent formation of polar molecules in cold atom systems means that it
is possible to generate customizable interactions with a similar form to those of the dressed Rydberg states \cite{micheli2007,pupillo2009}. Under certain circumstances, they might offer better interaction properties than cold Rydberg atoms. For example the form of the interaction can be controlled at very short distances, such that it may be possible to adjust the interaction with multiple avoided crossings so that the regions where the $1/r$ and dipole-dipole interaction curves match are longer \cite{micheli2007}. Particular benefits of polar molecules are narrow linewidths of the transitions that lead to long lived excited states and low scattering rates (which would lead to low heating). Dipoles and therefore interaction strengths are induced by external electric fields, which makes them highly controllable, although for very strong fields there could be issues with anisotropy. Since the form of the interactions is very similar to that for dressed Rydberg atoms, the arguments of the previous section regarding the interaction matrices should apply. We do not have access to sufficient information to make quantitative  predictions regarding the use of polar molecules at the current time.

\subsection{Feshbach resonances and cold atoms}

Interactions between cold atoms can be magnetically tuned using Feshbach resonances \cite{RevModPhys80885}, such that they have a delta function form, $4\pi \hbar^2 a(B) \delta(\rvec) / M$, where
\begin{equation}
a(B)=a_{bg}\left(1-\frac{\Delta_{F}}{B-B_0}\right)
\end{equation}
where $\Delta_{F}$ is the resonance width, $a_{bg}$ is the background scattering length, and $B$ the magnetic field \cite{RevModPhys80885,bartmettler2015}.

The ratio of the interaction matrices diverges for very small Wannier-like functions. However, we suggest that use of weak interactions induced by Feshbach resonances might be useful to correct the underestimation in the ratio of the interaction matrices on short distances when using dressed Rydberg states \footnote{For polar molecules, the same correction may not be possible if the Feshbach resonance is used to bind the molecule.}. For large $B$ fields, anisotropy in Rydberg states could become a problem if making this correction. 

\subsection{Cold ions}

The scaling that is required for the interactions between particles means
that although $1/r$ interaction is available for cold ions loaded into an optical
lattice, the energy scales are wrong for quantum
simulation of materials by several orders of magnitude. The interactions must be scaled by a
factor $m_ea_{\rm nn,CM}/Ma_{\rm nn}$ (as for the $-k_{e}e^2Z/r$ nuclear potential, there is a cancellation of factors of $a_{\rm nn}$ and $a_{\rm nn,CM}$ that emerge when scaling $r$). However, the interactions between ions are fixed by their charge, so cold ions in optical lattices will have poorly representative band structures. This is confirmed by Fig. \ref{fig:interactionscale}, where increasing the size of the potential by only a factor of 2 dramatically changes the band structure close to the Fermi surface.

\section{Operating temperature and decoherence}
\label{sec:coherence}

For the final section of this article, we will focus on decoherence effects in dressed Rydberg atoms. These effects can be divided into decoherence by atom heating due to the optical lattice and laser dressing, and decoherence by atom losses. 

\subsection{Heating due to the optical lattice}

A source of decoherence is the absorption of photons used to form the optical lattice and to dress Rydberg atoms. This problem has been treated in depth in \cite{PhysRevA.82.063605}. The optical lattice is painted or generated by holograms using a laser of wavelength $\lambda_L$ (detuning $\bar{\Delta}$) and Rabi frequency $\bar{\Omega}$, distinct from those involved in the dressing of the Rydberg states. Energy is absorbed by each atom at a rate $\dot{E}_{L} = \gamma_0 = \Gamma_{\rm eg} \left(\bar{\Omega}^2/4\bar{\Delta}^2\right) 2 E_R$, where  $E_R = (h/\lambda_{L})^2 /2M$ is the recoil energy, $\Gamma_{\rm eg}$ is the natural linewidth and $\lambda_{\rm eg}$ is the wavelength of the $^1S_0\rightarrow ^1P_1$ transition \cite{PhysRevA.82.063605}.  In the expression for $\dot{E}_{L}$, the factor of two appearing in front of the recoil energy given above corresponds to the average kinetic energy of an atom recoiling from absorbing a laser photon and emitting spontaneously into the full $4\pi$ steradians solid angle \cite{grimm2000}. The maximum lattice depth can be calculated by taking the leading term of the Maclaurin expansion of Eqn. \ref{eqn:extpot}, which can be computed using $\erf(z)\approx 2z/\sqrt{\pi}$ as $V^{(AT)}_{\rm ext}(0) = - a_{\rm nn,CM}m_{e}k_e e^2 Z (2^{3/2}/\sqrt{\pi}+D)/w a_{\rm nn}M$. This should be matched to the optical lattice parameters via $V_{\rm lattice}=\hbar\bar{\Omega}^2/4\bar{\Delta}$ which corresponds to the maximum laser intensity. $\bar{\Delta}=c(1/\lambda_{L}-1/\lambda_{eg})$, so the lattice depth fixes $\bar{\Omega}$. We select $\lambda_{L}=532$nm for all atomic species.
This estimated rate of heating due to the lattice with $D=1.24$ is shown in Tab. \ref{tab:draparams}.

\subsection{Heating due to laser dressing}

	Heating associated with the laser dressing is more rapid because the scheme only works when the lasers are tuned fairly close to the atomic transitions. The first requirement that we must satisfy is that the dressing is strong enough. The strength of the Rydberg-Rydberg dressed interaction saturates at $\approx \Omega\alpha^3$. As discussed in Sec. \ref{sec:rydberg}, since we are driving a $4s^2$ to $4sn_{\rm Ryd}s$ in $^{43}$Ca or $5s^2$ to $5sn_{\rm Ryd}s$ transition in $^{87}$Sr, we need a two-photon laser excitation scheme, which means excitation via an intermediate state. The two lasers have Rabi-frequencies of $\omega_i$ and detuning $\delta_i$, i=1,2. We want to achieve a two-photon Rabi frequency of $\Omega_{\mathrm{2ph}} = \omega_1\omega_2/2\delta_1$ and a two-photon detuning of $ \Delta_{\mathrm{2ph}} = \delta_1+\delta_2$. In this scheme, heating will arise from driving the lower transition, whereas driving the upper transition will lead to losses from the Rydberg state. This occurs because the Rydberg states decay via a cascade, whereas the 4s4p state decays back to the $4s^2$ ground state. The heating rate derived using similar arguments to Refs. \cite{PhysRevA.82.063605,grimm2000} is $\dot{E}_{\rm Ryd} = (h^2/\lambda_{^3P_1}^2 M)(\Gamma_{^3P_1}\omega_{1}^2/4\delta_{1}^2)$, the first factor being the kinetic energy of the recoiling atom, the second factor is the scattering rate. If we pick $\omega_1=\omega_2$, then $\dot{E}_{\rm Ryd} = (h^2/\lambda_{^3P_1}^2 M)(\Gamma_{^3P_1}\Omega_{\textrm{2ph}}/2\delta_{1})$. It is because of the heating due to the dressing scheme that we have chosen to study the Alkaline-Earth elements, rather than the Alkali metals common in cold atom experiments. In the case of the Alkali atoms, the intermediate states would be the $nP$ states closest to the ground state, all of which have $\Gamma \sim 2\pi \times 6\;\mathrm{MHz}$ linewidths and correspondingly high scattering rates. The Alkaline-earth metals Ca and Sr, however, benefit from a low-lying state with linewidth, $\Gamma_{^3P_1}$, on the order of $2\pi\times 7\mathrm{kHz}$, which is narrow enough to allow laser cooling to very close to quantum degeneracy.

Heating rates due to dressing are summarized in Table \ref{tab:draparams}.  For $^{43}$Ca, calculations assume driving a two photon transition exploiting the 657 nm $\mathrm{^1S_0}\rightarrow \; \mathrm{^3P_1}$ intercombination transition ($\Gamma_{^3P_1}/2\pi=6 \;\mathrm{kHz}$). For $^{87}$Sr, this transition has $\lambda_{^3P_1}=689$ nm and linewidth $\Gamma_{^3P_1}/2\pi = 7.5\;\mathrm{kHz}$.  To suppress the heating rate requires detuning the lasers very far from the one-photon resonances, with a corresponding increase in Rabi frequency. The smaller values of $\omega_1$ and $\delta_1$ that we have chosen in this example are challenging, requiring large laser intensity (around 130 - 200 mW when focused into $50\times50\;\mu\textrm{m}^2$). We compute the laser power required by noting that in terms of the laser intensity $I_0$, $\omega_1 = \Gamma\sqrt{I_0/2I_{sat}}$, where the saturation intensity is $I_{sat}=2\pi^2\hbar\Gamma_{^3P_1}\c/3\lambda_{^3P_1}^3$ \cite{louden1997}. The table shows that heating rates are on the order of a few nK / ms, slightly larger than the heating from the lattice.

We estimate that it should be possible to reduce heating further using Mg. We do not have access to the reliable data regarding quantum defects for Mg Rydberg levels, but we make the observation that given the similarity between the Ca and Sr Rydberg levels, a suitable Rydberg level, giving useful interaction parameters, will almost certainly exist for Mg. The Mg intercombination line is even more highly suppressed than those of Ca and Sr, because Mg does not have the strong spin-orbit coupling of the heavier elements. Hence, the Mg $^1S_1\to^3P_1$ transition linewidth is two orders of magnitude lower, $\Gamma_{^3P_1} = 2\pi\times 31$ Hz; for this transition, $\lambda_{^3P_1}=457$ nm, so while the recoil energy is 3.55 times larger than for $^{43}$Ca, the heating rate is significantly reduced. With such a small linewidth, and correspondingly small saturation intensity, the laser power required to drive the intercombination transition is significantly increased, but with the advantage of a slower heating rate: see Table \ref{tab:draparams}. We calculate that about 1.6 W of laser power focused into an area of $25\times25\;\mu\textrm{m}^2$ is sufficient to drive the transition.

\subsection{Atom losses}

Next, we discuss atom losses. The main loss will be due to the finite Rydberg atom lifetime $\sim 21 \;\mathrm{\mu sec}$, which, since the probability of exciting the Rydberg state is $\alpha^2$, can be expected to be enhanced by laser dressing, by a factor of $\alpha^{-2}$, (recall that the amplitude of the Rydberg state is $\alpha=\Omega_{\mathrm{2ph}}/2\Delta_{\mathrm{2ph}})$.) The lifetime of the Sr 5s20s state is about 3.1 $\mu$s \cite{millen2011}, and using the $n_{\rm Ryd}^3$ scaling, we estimate that the Sr 5s38s state lifetime will be 21 $\mu$s (we expect this lifetime will be similar in other alkaline earths, which are estimated in the same way). For our parameters, $\alpha^{-2}\gtrsim 100$. This limits the experiments to around 1 to 4 ms. Although this is very short, the lasers can be tuned to improve this situation. Finally, we note that the dressed atom schemes are subject also to collective decoherence effects, which are the subject of experimental and theoretical work \cite{glaetzle2015,zeiher2016,fossfeig2013}. So far, the results of experiment of Ref. \cite{zeiher2016} achieved improved Rydberg lifetimes, however the increase was far below the theoretical maximum.

\subsection{Consequences of decoherence effects}

For graphene, $t\sim 3$eV, so the hopping energy in the quantum simulator is approximately $t_{\rm AT}=m_{e} a_{\rm nn,CM}^2 3\mathrm{eV}/hMa_{\rm nn}^2$. Typical timescales to hop between sites are $\hbar/2t_{\rm AT}$ on a tight binding chain (see e.g. \cite{siber2006}) which corresponds well to the group velocity at the Fermi surface of a half filled chain. Using the group velocity at the Fermi surface of graphene, which is $v_{g}=3ta/2\hbar$, we find that the typical timescales, $\tau$, to hop between carbon atoms are slightly slower $\tau = 2\hbar/3t_{\rm AT}$.

The bandwidth of the $\pi$ bands in graphene is $6t\approx 18 \;\mathrm{eV}$. Rating the quality of the quantum simulator depends on how well resolved the bandstructure needs to be. To resolve bandstructure features to 10\%, corresponds to an energy scale of $m_e a_{\rm nn,CM}^2/M a_{\rm nn}^2 \times 1.8\mathrm{eV}$ in the cold atom system. The temperature of the cold atom system needs to be low enough to resolve these details, and can be estimated as $T_{10\%}=m_e a_{\rm nn,CM}^2/M a_{\rm nn}^2 \times 1.8\mathrm{eV}/k_{B}$, where $k_B$ is Boltzman's constant. Finally, the temperatures must remain stable for long enough for hopping between lattice sites. Assuming that the initial temperature of the simulator is much lower than $T_{10\%}$, then a rough estimate of the number of hops until the simulator is too warm to resolve bandstructure features to 10\% is $T_{10\%}/\tau \dot{E}$, where $\dot{E}$ is the total heating rate. This estimate will be better for simulators with higher operating temperatures (and larger numbers of hops per experiment). The number of hops may also be limited by Rydberg lifetimes as $\tau_{\rm Ryd}/\alpha^2\tau$. The approximate number of hops per experiment taking the smallest of the two values is shown in Table \ref{tab:draparams}. This indicates that so long as linewidths remain small, lighter atoms and shorter lattices perform better (have more hops) before heating becomes an issue, and that detuning should be maximized.  Many condensed matter processes require only hops to and from a neighboring site, so a few hops should be sufficient to bring the simulator into equilibrium in many cases, especially if the interaction is turned on adiabatically. As an extreme example, just 3 hops are required to restore equilibrium to a Bose-Hubbard model undergoing a quench from a superfluid to Mott insulator state \cite{kollath2007a}. Our estimates indicate that Mg DRAs should be used.

\section{Conclusions}
\label{sec:conclusions}

We have examined some of the challenges involved in making a quantum simulator that can emulate multi-band materials. Optical lattices are fixed and therefore naturally follow the  Born-Oppenheimer approximation where phonons are neglected. Within Born--Oppenheimer, the full many-body Hamiltonian contains three terms: Kinetic energy of the electrons, interaction between electrons and the lattice and the interaction between electrons. Since the kinetic energies of electrons in a condensed matter system and cold atoms in a quantum simulator are the same, we focused on ways in which the two interaction potentials could be approximated in a multiband quantum simulator.
\begin{enumerate}
\item Our first aim was understanding how optical lattice resolution changes the bandstructure of a multi-band quantum simulator. We have performed DFT calculations to examine these effects. This is a step towards determining the forms of optical lattice suitable for quantum simulators with the approximate band structures of real materials. Our primary conclusions are that:
\begin{itemize}
\item Resolution effects do not modify the band structure around the Fermi surface over a wide range of beam waist sizes.
\item Core state energies may overlap valence states as $w/a_{\rm nn}$ becomes large and the lattice potential becomes sinusoidal. Since core and valence states are typically well separated in real materials, standard sinusoidal optical lattices may not be suitable for many band applications. 
\item Optical resolution effects are minimized by making $a_{\rm nn}/w$ large. However $a_{\rm nn}$ cannot be increased indefinitely since operating temperatures decrease with $a_{\rm nn}$. This can be partially mitigated using short wavelength lasers and lighter atoms to increase operating temperatures. 
\end{itemize}
\item Our second aim was to identify electron analogues with a suitable magnitude of interaction matrices. Our second set of conclusions are as follows:
\begin{itemize}
\item DFT calculations indicate that bandstructures maintain qualitatively similar forms even if interaction strengths vary by around a factor of 2.
\item The strength of interactions between cold ions is fixed by the electron charge and is too strong for probing multiband physics.
\item Dressed Rydberg atoms have highly tuneable dipole-dipole interactions. Calculations using van Vleck perturbation theory show that interaction matrices are similar to those of a Coulomb potential over a range of Wannier function sizes. Quantum simulation is possible, but limited to a few hops.
\item Polar molecules are also highly tuneable. We have not made detailed calculations here, but they could offer a similar benefits to Rydberg atoms. 
\item Cold atoms interacting via Feshbach resonances have interaction matrix elements that diverge relative to those of the Coulomb interaction for Wannier functions with small radius.
\end{itemize}
\end{enumerate}

Our interpretation of these results is that simulation of multiband materials is possible using DRAs, but it will be approximate rather than predictive, and right at the limit of currently available Rydberg dressing and optical lattice technology. This paper discussed the toy systems of graphene and BN because of the need to make contact with the well understood electronic structure of these materials. The goal is that more complex systems could be simulated. Interactions would need tuning for each system to be investigated, and could not be selected {\it ab-initio}. Schemes following these principles could be used as test beds for numerical condensed matter techniques. Future work should focus on improving the interactions between the electron analogues while decreasing decoherence effects. For example, single phonon Rydberg dressing schemes with very low heating rates may be possible using atoms with a P ground state, such as In \cite{kim2009}. It would also be very interesting to find ways to introduce tuneable phonon effects into multiband quantum simulators.

\bibliographystyle{unsrt}
\bibliography{bestquantsim_final_rev3}

\end{document}